\documentclass[10pt,twocolumn,superscriptaddress,floats,showpacs,nobalancelastpage,longbibliography,prb]{revtex4-2}

\usepackage{array,longtable}
\newcolumntype{L}{>{\tiny $}p{0.33\columnwidth}<{$}}
\newcolumntype{M}{>{\scriptsize $}p{0.33\columnwidth}<{$}}
\newcolumntype{N}{>{\scriptsize $}p{0.43\columnwidth}<{$}}
\usepackage{amsmath}
\usepackage{amssymb}
\usepackage{amsfonts}
\usepackage{graphicx}
\usepackage{hyperref}
\usepackage{tabularx}

\usepackage{float}
\usepackage{graphics}
\usepackage[caption=false]{subfig}
\usepackage{tikz}
\usepackage{times}

\allowdisplaybreaks[1]

\newif\ifhyper
\hypertrue
\ifhyper
\hypersetup{
   citecolor = {green},
   colorlinks = {true}, % false
   urlcolor = {blue}, % magenta
   linkcolor = {blue}
}
\fi

\newenvironment{diagram}
{
\begin{tikzpicture}[baseline = (X.base),every node/.style={scale=0.8},scale=0.45]
}
{
\end{tikzpicture}
}

\begin{document}

\title{Generating Function for Tensor Network Diagrammatic Summation}

\author{Wei-Lin Tu}
\affiliation{Institute for Solid State Physics, University of Tokyo, Kashiwa, Chiba 277-8581, Japan}
\affiliation{Division of Display and Semiconductor Physics, Korea University, Sejong 30019, Korea}

\author{Huan-Kuang Wu}
\affiliation{Department of Physics, Condensed Matter Theory Center and Joint Quantum Institute, University of Maryland, College Park, Maryland 20742, USA}

\author{Norbert Schuch}
\affiliation{Max-Planck-Institut f\"ur Quantenoptik, Hans-Kopfermann-Stra{\ss}e 1, 85748 Garching, Germany}
\affiliation{Munich Center for Quantum Science and Technology, Schellingstra{\ss}e 4, 80799 M{\"u}nchen, Germany}
\affiliation{University of Vienna, Department of Physics, Boltzmanngasse 5, 1090 Wien, Austria}
\affiliation{University of Vienna, Department of Mathematics, Oskar-Morgenstern-Platz 1, 1090 Wien, Austria}

\author{Naoki Kawashima}
\affiliation{Institute for Solid State Physics, University of Tokyo, Kashiwa, Chiba 277-8581, Japan}

\author{Ji-Yao Chen}
\email{Ji-Yao.Chen@mpq.mpg.de}
\affiliation{Max-Planck-Institut f\"ur Quantenoptik, Hans-Kopfermann-Stra{\ss}e 1, 85748 Garching, Germany}
\affiliation{Munich Center for Quantum Science and Technology, Schellingstra{\ss}e 4, 80799 M{\"u}nchen, Germany}

\date{\today}

\begin{abstract}

The understanding of complex quantum many-body systems has been vastly boosted by tensor network (TN) methods. Among others, excitation spectrum and long-range interacting systems can be studied using TNs, where one however confronts the intricate summation over an extensive number of tensor diagrams. Here, we introduce a set of generating functions, which encode the diagrammatic summations as leading order series expansion coefficients. Combined with automatic differentiation, the generating function allows us to solve the problem of TN diagrammatic summation. We illustrate this scheme by computing variational excited states and dynamical structure factor of a quantum spin chain, and further investigating entanglement properties of excited states. Extensions to infinite size systems and higher dimension are outlined.

\end{abstract}

%\pacs{}

\maketitle

\section{Introduction}

The study of quantum many-body systems using tensor networks (TNs) has witnessed great success in the last three decades~\cite{Schollwock2011, Orus2019, Cirac2020}. Originally, TN methods were developed to efficiently capture ground state properties of many-body lattice models with short-range interaction~\cite{White1992, Ostlund1995, Verstraete2004b, Vidal2008}. Later on, numerous progress has been made in various directions, including determining low-energy excited states~\cite{Vanderstraeten2019b}, exploring dynamical and finite temperature properties~\cite{Paeckel2019}, and finding valuable applications in long-range interacting systems~\cite{Stoudenmire2017, Motta2020}. These developments not only deepen our theoretical understanding of many-body systems~\cite{Haegeman2013}, but also bridge TN methods to real experiments~\cite{Mena2020}.

New developments also bring challenges, however. Both quasiparticle excited states~\cite{Ostlund1995, Pirvu2012a,Vanderstraeten2019a, Ponsioen2020} and global observables contain contributions with a sizable number of tensor diagrams, due to the fact that quasiparticle or local operators of global observables can be on arbitrary patch of the lattice. Except in a few cases where efficient summation techniques have been proposed~\cite{Vanderstraeten2015, Corboz2016, McCulloch2007, Michel2010}, most notably the matrix product operator (MPO) representation of global observables~\cite{McCulloch2007}, extensive and costly tensor diagram manipulation seems to be unavoidable and becomes the bottleneck in modern TN applications. Thus, an efficient and universal approach for TN diagrammatic summation is highly called for.

Another domain where diagrammatic summations frequently appear is the perturbation theory of interacting quantum fields~\cite{Peskin1995, Altland2010}. There, for correlation functions containing summations of Feynman diagrams, one can introduce a source field and formally represent correlation functions as derivatives of the perturbed partition function, known as the generating functional method~\cite{Peskin1995}. Given the close relation between the trio of TN methods, many-body systems, and quantum field theory (QFT), and the pictorial similarity of tensor diagram and Feynman diagram, it is tempting to look for a generating function formalism in TNs, where certain derivatives can compactly represent the summations of TNs. This is plausible, as partition functions of classical statistical models are known to be representable as TNs~\cite{Haegeman2016}.

In this work, inspired by the generating functional method in QFT, we propose a set of generating functions for TNs, which encode TN diagrammatic summations as leading order expansion coefficients. It then requires taking derivatives of the generating functions, which can be accomplished with automatic differentiation (AD)~\cite{Bartholomew2000,Baydin2018,Liao2019}. To illustrate the scheme, we investigate the low-lying spectrum of a quantum spin chain with periodic uniform matrix product state (MPS) and the excitation ansatz~\cite{Pirvu2012a}, and subsequently study entanglement properties of excited states which, to our knowledge, were rarely studied due to the overwhelmingly large number of tensor diagrams involved.

\section{Periodic uniform MPS and excitations}

Let us consider a translationally invariant quantum spin chain with $N$ sites. For simplicity, we assume the ground state is unique, and can be approximated by a periodic uniform MPS with the form:
\begin{equation}
|\Psi(A)\rangle = 
\begin{diagram}
\draw (0.5, 1.5) .. controls (0, 1.5) and (0, 2) .. (0.5, 2);
\draw (0.5, 1.5) -- (1, 1.5); \draw[] (1, 2) rectangle (2, 1); \draw (1.5, 1.5) node {$A$};
\draw (2, 1.5) -- (3, 1.5); \draw[] (3, 2) rectangle (4, 1); \draw (3.5, 1.5) node {$A$};
\draw (4, 1.5) -- (5, 1.5); \draw[] (5, 2) rectangle (6, 1); \draw (5.5, 1.5) node {$A$}; \draw (6, 1.5) -- (6.5, 1.5); 
\draw (7, 1.5) node {$\ldots$};
\draw (7.5, 1.5) -- (8, 1.5); \draw[] (8, 2) rectangle (9, 1); \draw (8.5, 1.5) node {$A$};
\draw (9, 1.5) -- (10, 1.5); \draw[] (10, 2) rectangle (11, 1); \draw (10.5, 1.5) node {$A$};
\draw (11, 1.5) -- (11.5, 1.5);
\draw (11.5, 1.5) .. controls (12, 1.5) and (12, 2) .. (11.5, 2);
\draw (1.5, 1.3) node (X) {};
\draw (1.5, 1) -- (1.5, 0.5); \draw (1.5, 0) node {$s_1$};
\draw (3.5, 1) -- (3.5, 0.5); \draw (3.5, 0) node {$s_2$};
\draw (5.5, 1) -- (5.5, 0.5); \draw (5.5, 0) node {$s_3$};
\draw (7, 0) node {$\ldots$};
\draw (8.5, 1) -- (8.5, 0.5); \draw (8.5, 0) node {$s_{N-1}$};
\draw (10.5, 1) -- (10.5, 0.5); \draw (10.5, 0) node {$s_N$};
\end{diagram},
\label{eq:uniformMPS}
\end{equation}
where the same rank-3 tensor $A$ with dimension $d\times D\times D$ is repeated on every site, and the left boundary is contracted with the right boundary. Here $s_i=1,\ldots,d$ represents basis of $d$-dimensional local Hilbert space, while $D$ is the virtual bond dimension, controlling the accuracy of the MPS ansatz. We further denote the one-site translation operator as $\hat{T}$ with $\hat{T}|s_1,s_2,\ldots, s_N\rangle=|s_N,s_1,\ldots, s_{N-1}\rangle$, which satisfies $\hat{T}^N=1$. By construction, $|\Psi(A)\rangle$ is translationally invariant with momentum $k=0$, i.e., $\hat{T}|\Psi(A)\rangle=|\Psi(A)\rangle$. For a given model, the ground state tensor $A$ can be optimized 
using the conjugate gradient method, with gradient obtained from AD of the computation graph for energy~\cite{Liao2019}~\footnote{Variational uniform MPS algorithm for infinite size systems~\cite{Zauner2018} can provide a good initial tensor.}.

With ground state tensor at hand, excited states can be constructed using the single mode approximation~\cite{Auerbach1994}, which correspond to one-particle excitation and work well for a broad range of models~\cite{Haegeman2012, Zou2018, Vanderstraeten2018}. In full generality, one can perturb the ground state by replacing one site tensor $A$ with a new tensor $B$ which is yet to be determined, and then build up a Bloch state using translation operator~\cite{Ostlund1995, Pirvu2012a}, taking the form:
\begin{equation}
|\Phi_k(B)\rangle=\sum_{j=0}^{N-1}\mathrm{e}^{-ikj}\hat{T}^j
\begin{diagram}
\draw (0.5, 1.5) .. controls (0, 1.5) and (0, 2) .. (0.5, 2);
\draw (0.5, 1.5) -- (1, 1.5); \draw[] (1, 2) rectangle (2, 1); \draw (1.5, 1.5) node {$B$};
\draw (2, 1.5) -- (3, 1.5); \draw[] (3, 2) rectangle (4, 1); \draw (3.5, 1.5) node {$A$}; \draw (4, 1.5) -- (4.5, 1.5); 
\draw (5, 1.5) node {$\ldots$};
\draw (5.5, 1.5) -- (6, 1.5); \draw[] (6, 2) rectangle (7, 1); \draw (6.5, 1.5) node {$A$};
\draw (7, 1.5) -- (7.5, 1.5);
\draw (7.5, 1.5) .. controls (8, 1.5) and (8, 2) .. (7.5, 2);
\draw (1.5, 1.3) node (X) {};
\draw (1.5, 1) -- (1.5, 0.5); \draw (1.5, 0) node {$s_1$};
\draw (3.5, 1) -- (3.5, 0.5); \draw (3.5, 0) node {$s_2$};
\draw (5, 0) node {$\ldots$};
\draw (6.5, 1) -- (6.5, 0.5); \draw (6.5, 0) node {$s_{N}$};
\end{diagram},
\label{eq:excitation}
\end{equation}
where tensor $B$ contains variational parameters for excited state. $|\Phi_k(B)\rangle$ then is an eigenstate of translation operator with eigenvalue $\mathrm{e}^{ik}$, where momentum $k=2\pi m/N, m=0,1,\ldots,N-1$. Due to momentum superposition, a summation of $N$ different tensor diagrams appears in Eq.~\eqref{eq:excitation}, which will be our main focus. Since $|\Phi_k(B)\rangle$ depends on tensor $B$ linearly, variationally optimizing $B$ boils down to a generalized eigenvalue problem: $\mathbf{H}_{\mu\nu}\mathbf{B^{\nu}} = E\mathbf{N}_{\mu\nu}\mathbf{B}^{\nu}$, where $E$ is the generalized eigenvalue, and $\mathbf{H}$ $(\mathbf{N})$ is the effective Hamiltonian (norm) matrix in the variational space, with $\mathbf{H}_{\mu\nu}=\frac{\partial^2 }{\partial\overline{\mathbf{B}}^{\mu}\partial \mathbf{B}^{\nu}}\langle\Phi_k(B)|\hat{H}|\Phi_k(B)\rangle$, $\mathbf{N}_{\mu\nu}=\frac{\partial^2 }{\partial\overline{\mathbf{B}}^{\mu}\partial \mathbf{B}^{\nu}}\langle\Phi_k(B)|\Phi_k(B)\rangle$. Here $\overline{B}$ is complex conjugate of $B$, whose component after vectorization is denoted as $\mathbf{B}^{\nu}$. Since momentum is a good quantum number, we have suppressed the dependence of $\mathbf{H}, \mathbf{N}, E$, and $\mathbf{B}$ on momentum $k$. Solving the generalized eigenvalue equation in each momentum sector, one recovers the low-energy spectrum~\footnote{Due to gauge degree of freedom and orthogonality to ground state, in each momentum sector, one can obtain $(d-1)D^2$ valid solutions for excited states~\cite{Haegeman2012}.}.

To construct $\mathbf{H}$ and $\mathbf{N}$, one needs to sum over $N$ different tensor diagrams for each, with MPO representation of the Hamiltonian $\hat{H}$. These extensive tensor diagram summations are the main obstacles of computing the excitation ansatz, rendering manipulating excited states unfavorable. Below we introduce our formalism based on simple yet powerful generating functions with the following strategy: to compute $\mathbf{H}$ or $\mathbf{N}$, we will first construct a suitable generating function, and then use AD to compute the derivative~\cite{Liao2019}, which will reproduce $\mathbf{H}$ or $\mathbf{N}$ and is much simpler than directly summing all diagrams. In this way, we will get rid of all the tensor diagram summations, making it possible to investigate detailed properties of excited states. Note that, unlike generating functionals in QFT, whose closed-form expressions are rare, the TN generating functions and their derivatives can be computed numerically exact with AD. We find that, depending on the origins of diagrammatic summation, the generating functions can be divided into two classes, one for TN state and the other for TN operators, which we introduce separately.

\section{Generating function for state}

As shown in Eq.~\eqref{eq:excitation}, the extensive tensor diagrams only differ by the location of tensor $B$ and a position dependent phase factor. It is insightful to make the following observation: for a given tensor $B$, the excitation ansatz Eq.~\eqref{eq:excitation} can be expressed as: $|\Phi_k(B)\rangle = \frac{\partial}{\partial\lambda}|G_{\Phi}(\lambda)\rangle\Bigr |_{\lambda=0}$, with
\begin{equation}
|G_{\Phi}(\lambda)\rangle =
\begin{diagram}
\draw (0.5, 1.5) .. controls (0, 1.5) and (0, 2) .. (0.5, 2);
\draw (0.5, 1.5) -- (1, 1.5); \filldraw[fill={rgb, 255: red, 4; green, 128; blue, 255}] (1, 2) rectangle (2, 1); \draw (1.5, 1.3) node (X) {};
\draw (2, 1.5) -- (3, 1.5); \filldraw[fill={rgb, 255: red, 4; green, 128; blue, 255}] (3, 2) rectangle (4, 1);
\draw (4, 1.5) -- (5, 1.5); \filldraw[fill={rgb, 255: red, 4; green, 128; blue, 255}] (5, 2) rectangle (6, 1);
\draw (6, 1.5) -- (6.5, 1.5);
\draw (7, 1.5) node {$\ldots$};
\draw (7.5, 1.5) -- (8, 1.5); \filldraw[fill={rgb, 255: red, 4; green, 128; blue, 255}] (8, 2) rectangle (9, 1);
\draw (9, 1.5) -- (10, 1.5); \filldraw[fill={rgb, 255: red, 4; green, 128; blue, 255}] (10, 2) rectangle (11, 1);
\draw (11, 1.5) -- (11.5, 1.5);
\draw (11.5, 1.5) .. controls (12, 1.5) and (12, 2) .. (11.5, 2);
\draw (1.5, 1) -- (1.5, 0.5); \draw (1.5, 0) node {$s_1$};
\draw (3.5, 1) -- (3.5, 0.5); \draw (3.5, 0) node {$s_2$};
\draw (5.5, 1) -- (5.5, 0.5); \draw (5.5, 0) node {$s_3$};
\draw (7, 0) node {$\ldots$};
\draw (8.5, 1) -- (8.5, 0.5); \draw (8.5, 0) node {$s_{N-1}$};
\draw (10.5, 1) -- (10.5, 0.5); \draw (10.5, 0) node {$s_N$};
\end{diagram},
\label{eq:excitationGen}
\end{equation}
where the tensor on the $j$-th site in $|G_{\Phi}(\lambda)\rangle$ is given by $A_j(\lambda) = A+\lambda\mathrm{e}^{-ik(j-1)}B$, $\lambda\in\mathbb{R}$, represented by blue squares. Here, to simplify the notation, we have suppressed the dependence of $|G_{\Phi}(\lambda)\rangle$ on tensor $A, B$ and momentum $k$, keeping only $\lambda$-dependence explicitly. Expanding $|G_{\Phi}(\lambda)\rangle$ into power series of $\lambda$, we find that the ground (excited) state $|\Psi(A)\rangle$ $(|\Phi_k(B)\rangle)$ is contained in the zeroth (first) order term, both of which lie in the tangent space of the MPS manifold, while higher order terms are outside of the tangent space due to nonlinearity in tensor $B$~\cite{Haegeman2014}. Thus, we can eliminate the tensor diagram summation in $|\Phi_k(B)\rangle$ by computing the first order derivative of $|G_{\Phi}(\lambda)\rangle$. Interestingly, Eq.~\eqref{eq:excitationGen} bears a similarity with the generating functional in QFT, where the parameter $\lambda$ plays the role of source field in the latter. Note that, although $|\Psi(A)\rangle$ and $|\Phi_k(B)\rangle$ are translationally invariant, the generating function $|G_{\Phi}(\lambda)\rangle$ is in general not invariant under one-site translation, except at momentum $k=0$.

With $|G_{\Phi}(\lambda)\rangle$, the norm square of excited state can be expressed as $|||\Phi\rangle||^2\equiv\langle\Phi_{k}(B)|\Phi_{k}(B)\rangle=\frac{\partial^2}{\partial\lambda'\partial\lambda}\langle G_{\Phi}(\lambda')|G_{\Phi}(\lambda)\rangle\Bigr |_{\lambda'=\lambda=0}$. Using translation invariance of $|\Phi_{k}(B)\rangle$, we can lower the order of derivative with the following generating function for the excited state norm:
\begin{equation}
G_{||\Phi||}(\lambda)=
\begin{diagram}
\draw (0.5, 1.5) .. controls (0, 1.5) and (0, 2) .. (0.5, 2);
\draw (0.5, 1.5) -- (1, 1.5); \filldraw[fill={rgb, 255: red, 4; green, 128; blue, 255}] (1, 2) rectangle (2, 1);
\draw (2, 1.5) -- (3, 1.5); \filldraw[fill={rgb, 255: red, 4; green, 128; blue, 255}] (3, 2) rectangle (4, 1);
\draw (4, 1.5) -- (5, 1.5); \filldraw[fill={rgb, 255: red, 4; green, 128; blue, 255}] (5, 2) rectangle (6, 1);
\draw (6, 1.5) -- (6.5, 1.5);
\draw (7, 1.5) node {$\ldots$};
\draw (7.5, 1.5) -- (8, 1.5); \filldraw[fill={rgb, 255: red, 4; green, 128; blue, 255}] (8, 2) rectangle (9, 1);
\draw (9, 1.5) -- (10, 1.5); \filldraw[fill={rgb, 255: red, 4; green, 128; blue, 255}] (10, 2) rectangle (11, 1);
\draw (11, 1.5) -- (11.5, 1.5);
\draw (11.5, 1.5) .. controls (12, 1.5) and (12, 2) .. (11.5, 2);
\draw (1.5, 1) -- (1.5, 0); \draw (3.5, 1) -- (3.5, 0); \draw (5.5, 1) -- (5.5, 0);
\draw (8.5, 1) -- (8.5, 0); \draw (10.5, 1) -- (10.5, 0);
\draw (1.5, 0.3) node (X) {};
\draw (0.5, -0.5) .. controls (0, -0.5) and (0, -1) .. (0.5, -1);
\draw (0.5, -0.5) -- (1, -0.5); \draw[] (1, 0) rectangle (2, -1); \draw (1.5, -0.5) node {$\overline{B}$};
\draw (2, -0.5) -- (3, -0.5); \draw[] (3, 0) rectangle (4, -1); \draw (3.5, -0.5) node {$\overline{A}$};
\draw (4, -0.5) -- (5, -0.5); \draw[] (5, 0) rectangle (6, -1); \draw (5.5, -0.5) node {$\overline{A}$}; \draw (6, -0.5) -- (6.5, -0.5); 
\draw (7, -0.5) node {\ldots};
\draw (7.5, -0.5) -- (8, -0.5); \draw[] (8, 0) rectangle (9, -1); \draw (8.5, -0.5) node {$\overline{A}$};
\draw (9, -0.5) -- (10, -0.5); \draw[] (10, 0) rectangle (11, -1); \draw (10.5, -0.5) node {$\overline{A}$};
\draw (11, -0.5) -- (11.5, -0.5);
\draw (11.5, -0.5) .. controls (12, -0.5) and (12, -1) .. (11.5, -1);
\end{diagram},
\label{eq:normGen}
\end{equation}
with which, the norm square can be obtained as $|||\Phi\rangle||^2=N\frac{\partial}{\partial\lambda}G_{||\Phi||}(\lambda)\Bigr |_{\lambda=0}$. Here, the local tensor on site-$j$ of the ket layer is $A_j(\lambda)$, the same as appearing in Eq.~\eqref{eq:excitationGen}. 

Before proceeding further, let us discuss how to use generating functions in practice. Taking Eq.~\eqref{eq:normGen} as an example, we first compute $G_{||\Phi||}(\lambda=0)$ by contracting the network in a conventional manner with computational complexity $O(D^5)$, followed by a back-propagation using AD, with which the first order derivative at $\lambda=0$ is obtained automatically, hence the norm square~\cite{Liao2019}. Since the computational complexity of AD grows with the order of derivative, it is advisable to utilize a generating function with which low order derivative suffices.

With Eq.~\eqref{eq:normGen}, it is straightforward to find the generating function for the norm matrix $\mathbf{N}$:
\begin{equation}
G_{\mathbf{N}}(\lambda, B) = 
\begin{diagram}
\draw (0.5, 1.5) .. controls (0, 1.5) and (0, 2) .. (0.5, 2);
\draw (0.5, 1.5) -- (1, 1.5); \filldraw[fill={rgb, 255: red, 4; green, 128; blue, 255}] (1, 2) rectangle (2, 1);
\draw (2, 1.5) -- (3, 1.5); \filldraw[fill={rgb, 255: red, 4; green, 128; blue, 255}] (3, 2) rectangle (4, 1);
\draw (4, 1.5) -- (5, 1.5); \filldraw[fill={rgb, 255: red, 4; green, 128; blue, 255}] (5, 2) rectangle (6, 1);
\draw (6, 1.5) -- (6.5, 1.5); 
\draw (7, 1.5) node {$\ldots$};
\draw (7.5, 1.5) -- (8, 1.5); \filldraw[fill={rgb, 255: red, 4; green, 128; blue, 255}] (8, 2) rectangle (9, 1);
\draw (9, 1.5) -- (10, 1.5); \filldraw[fill={rgb, 255: red, 4; green, 128; blue, 255}] (10, 2) rectangle (11, 1);
\draw (11, 1.5) -- (11.5, 1.5);
\draw (11.5, 1.5) .. controls (12, 1.5) and (12, 2) .. (11.5, 2);
\draw (1.5, 1) -- (1.5, 0); \draw (3.5, 1) -- (3.5, 0); \draw (5.5, 1) -- (5.5, 0);
\draw (8.5, 1) -- (8.5, 0); \draw (10.5, 1) -- (10.5, 0);
\draw (1.5, 0.3) node (X) {};
\draw (0.5, -0.5) .. controls (0, -0.5) and (0, -1) .. (0.5, -1);
\draw (0.5, -0.5) -- (1, -0.5);
\draw (2, -0.5) -- (3, -0.5); \draw[] (3, 0) rectangle (4, -1); \draw (3.5, -0.5) node {$\overline{A}$};
\draw (4, -0.5) -- (5, -0.5); \draw[] (5, 0) rectangle (6, -1); \draw (5.5, -0.5) node {$\overline{A}$}; \draw (6, -0.5) -- (6.5, -0.5); 
\draw (7, -0.5) node {$\ldots$};
\draw (7.5, -0.5) -- (8, -0.5); \draw[] (8, 0) rectangle (9, -1); \draw (8.5, -0.5) node {$\overline{A}$};
\draw (9, -0.5) -- (10, -0.5); \draw[] (10, 0) rectangle (11, -1); \draw (10.5, -0.5) node {$\overline{A}$};
\draw (11, -0.5) -- (11.5, -0.5);
\draw (11.5, -0.5) .. controls (12, -0.5) and (12, -1) .. (11.5, -1);
\end{diagram},
\label{eq:normMatGen}
\end{equation}
where we simply omit tensor $\overline{B}$ in the bra layer of $G_{||\Phi||}(\lambda)$. From Eq.~\eqref{eq:normMatGen}, the norm matrix $\mathbf{N}$ can be obtained as $\mathbf{N} = N\frac{\partial }{\partial B}G_{\mathbf{N}}(\lambda, B)\Bigr |_{\lambda=1, B=0}$. Unlike the scalar $G_{||\Phi||}(\lambda)$ where the derivative is taken with respect to $\lambda$ at $\lambda=0$, $G_{\mathbf{N}}(\lambda, B)$ is a vector and the derivative is taken with respect to $B$ at $\lambda=1, B=0$, which can also be computed with AD~\cite{Paszke2017}.

\section{Generating function for operator}

Another type of TN diagrammatic summation originates from global observables. E.g., when computing structure factor, one needs to take into account correlations of local operator at all distances. Moreover, when applying TN methods to long-range interacting models which appear naturally in Rydberg quantum gas and quantum chemistry systems, efficient and faithful encoding of the long-range interaction is one of the main issues~\cite{Szalay2015}. Although approaches based on MPO and projected entangled-pair operator (PEPO) have been proposed, the bond dimension can be quite large, making them less appealing~\cite{Pirvu2010, ORourke2018, Li2019, ORourke2020}. Here we provide generating functions for global observables, without involving MPO whenever possible.

Let us first consider spin structure factor. Here, the summations arise from the Fourier transform of onsite spin operator $\hat{S}_j^{\alpha}$ (index $\alpha$ is in operator space):
$\hat{S}_k^{\alpha}=\frac{1}{\sqrt{N}}\sum_{j=1}^{N}\mathrm{e}^{-ikj}\hat{S}_j^{\alpha}$. A generating function can be introduced as follows:
\begin{equation}
\hat{G}_{S^\alpha}(\lambda)=
\begin{diagram}
\draw (1.5, 1) -- (1.5, 0.5); \filldraw[fill={rgb, 255: red, 255; green, 38; blue, 0}] (1.5, 0) circle (0.5); \draw (1.5, -0.5) -- (1.5, -1);
\draw (3.5, 1) -- (3.5, 0.5); \filldraw[fill={rgb, 255: red, 255; green, 38; blue, 0}] (3.5, 0) circle (0.5); \draw (3.5, -0.5) -- (3.5, -1);
\draw (5.5, 1) -- (5.5, 0.5); \filldraw[fill={rgb, 255: red, 255; green, 38; blue, 0}] (5.5, 0) circle (0.5); \draw (5.5, -0.5) -- (5.5, -1);
\draw (8.5, 1) -- (8.5, 0.5); \filldraw[fill={rgb, 255: red, 255; green, 38; blue, 0}] (8.5, 0) circle (0.5); \draw (8.5, -0.5) -- (8.5, -1); 
\draw (10.5, 1) -- (10.5, 0.5); \filldraw[fill={rgb, 255: red, 255; green, 38; blue, 0}] (10.5, 0) circle (0.5); \draw (10.5, -0.5) -- (10.5, -1);
\draw (1.5, -0.2) node (X) {};
\draw (7, 0) node {$\ldots$};
\draw (1.5, 1.5) node {$s_1$};
\draw (3.5, 1.5) node {$s_2$};
\draw (5.5, 1.5) node {$s_3$};
\draw (8.5, 1.5) node {$s_{N-1}$};
\draw (10.5, 1.5) node {$s_N$};
\draw (7, 1.5) node {$\ldots$};
\end{diagram},
\label{eq:sfGen}
\end{equation}
which is a tensor product of onsite operators, with $\hat{S}^{\alpha}_j(\lambda)=\mathbb{I}+\lambda\mathrm{e}^{-ikj}\hat{S}_j^{\alpha}$, $\lambda\in\mathbb{R}$, acting on the $j$-th site. 
One can then find $\hat{S}_k^{\alpha}=\frac{1}{\sqrt{N}}\frac{d}{d\lambda}\hat{G}_{S^\alpha}(\lambda)\Bigr |_{\lambda=0}$. Taking two layers of Eq.~\eqref{eq:sfGen} with independent parameters $\lambda$, $\lambda'$, followed by a second order derivative with respect to $\lambda$ and $\lambda'$, one arrives at the static structure factor. Higher order moments of local operator can be similarly obtained. Note that, with translation symmetry, one can combine $\hat{G}_{S^\alpha}(\lambda)$ and local spin operator to reduce the order of derivative.

Interestingly, by removing the $\lambda$-dependence in the first operator, and replacing the phase factor in the others with corresponding distance-dependent coefficient, Eq.~\eqref{eq:sfGen} can be used to efficiently encode long-range power-law decaying interactions. Moreover, this approach can be easily generalized to higher dimension, where PEPO technique for long-range interactions is imperfect~\cite{ORourke2018}. A simple counting suggests the number of terms for expectation value of long-range interaction is $N^2$, while with generating function Eq.~\eqref{eq:sfGen}, it is only of order $N$, where each term differs by the position of the first operator. Exploiting translation symmetry, the latter case can be further reduced to a single term. Without translation symmetry, a combination of Eq.~\eqref{eq:sfGen} and MPO technique is possible, which may lower the bond dimension of MPO.

While Eq.~\eqref{eq:sfGen} relies on the fact that $\hat{S}^{\alpha}$ is an onsite operator, the scheme can be generalized to the case where the local operator has support on more than one site, e.g., local Hamiltonian operator. Similar to Trotter gates, we introduce the following generating function for Hamiltonian operator under periodic boundary condition $\hat{H} = \sum_{j=1}^{N}\hat{h}_{j, j+1}$ (assuming $N$ even and nearest-neighbor interaction):
\begin{equation}
\hat{G}_H(\lambda) = 
\begin{diagram}
\draw (1.5, 1) -- (1.5, 0.5); \draw (3.5, 1) -- (3.5, 0.5); 
\filldraw[rounded corners, fill={rgb, 255: red, 255; green, 38; blue, 0}] (1, 0.5) rectangle (4, -0.5); 
\draw (1.5, -0.5) -- (1.5, -1); \draw (3.5, -0.5) -- (3.5, -1);
\draw (5.5, 1) -- (5.5, 0.5); 
\begin{scope}
\clip (4, 1) rectangle (6.5, -0.5);
\filldraw[rounded corners, fill={rgb, 255: red, 255; green, 38; blue, 0}] (5, 0.5) rectangle (8, -0.5); 
\end{scope}
\draw (5.5, -0.5) -- (5.5, -1);
\draw (7, 0) node {$\ldots$};
\draw (8.5, 1) -- (8.5, 0.5); \draw (10.5, 1) -- (10.5, 0.5); 
\filldraw[rounded corners, fill={rgb, 255: red, 255; green, 38; blue, 0}] (8, 0.5) rectangle (11, -0.5); 
\draw (8.5, -0.5) -- (8.5, -1); \draw (10.5, -0.5) -- (10.5, -1);
\draw (1.5, -0.95) node (X) {};
\draw (1.5, 1.5) node {$s_1$};
\draw (3.5, 1.5) node {$s_2$};
\draw (5.5, 1.5) node {$s_3$};
\draw (8.5, 1.5) node {$s_{N-1}$};
\draw (10.5, 1.5) node {$s_N$};
\begin{scope}
\clip (0.5, -0.9) rectangle (2.1, -2.6);
\filldraw[rounded corners, fill={rgb, 255: red, 255; green, 38; blue, 0}] (-1, -1) rectangle (2, -2);
\end{scope}
\draw (1.5, -2) -- (1.5, -2.5);
\filldraw[rounded corners, fill={rgb, 255: red, 255; green, 38; blue, 0}] (3, -1) rectangle (6, -2); 
\draw (3.5, -2) -- (3.5, -2.5); \draw (5.5, -2) -- (5.5, -2.5);
\draw (7, -1.5) node {$\ldots$};
\begin{scope}
\clip (7.5, -0.9) rectangle (9.1, -2.6);
\filldraw[rounded corners, fill={rgb, 255: red, 255; green, 38; blue, 0}] (6, -1) rectangle (9, -2);
\end{scope}
\draw (8.5, -2) -- (8.5, -2.5);
\begin{scope}
\clip (9.9, -0.9) rectangle (11.5, -2.6);
\filldraw[rounded corners, fill={rgb, 255: red, 255; green, 38; blue, 0}] (10, -1) rectangle (13, -2);
\end{scope}
\draw (10.5, -2) -- (10.5, -2.5);
\end{diagram},
\label{eq:hamGen}
\end{equation}
where the operator acting on the $(j, j+1)$ bond is given by $\hat{h}_{j, j+1}(\lambda)=\mathbb{I}+\lambda \hat{h}_{j, j+1}$, $\lambda\in\mathbb{R}$, represented by a red rectangle. It is then straightforward to verify $\hat{H}=\frac{d}{d\lambda}\hat{G}_H(\lambda)\Bigr |_{\lambda=0}$. For the effective Hamiltonian in the generalized eigenvalue equation, one could take the generating function for norm matrix Eq.~\eqref{eq:normMatGen}, and insert a MPO representation of the Hamiltonian, or the generating function Eq.~\eqref{eq:hamGen} with an additional derivative. The latter could be advantageous if the MPO operator has a relatively large bond dimension~\footnote{We have numerically verified this with the transverse field Ising model.}. Removing the boundary term, $\hat{G}_H(\lambda)$ can also be applied for systems with open boundary. Furthermore, taking two layers of $\hat{G}_H(\lambda)$ one can efficiently compute the energy variance.

A close comparison between generating functions for TN state, operator, and partition function in QFT suggests that, for excited state, it is constructed by introducing source field to the ground state tensor, while for operators, identity operator is taken as the reference. This formalism also reminds us of Lie algebra, whose generators can be obtained by expanding group elements around the identity. In that case, the generating function has an exponential form. Indeed, generating functions for moments and cumulants have been introduced in Ref.~\cite{West2015}, which took exponential forms with possible Trotter error and a finite difference method was then employed. In contrast, our approach does not have Trotter error, nor finite difference error, and the generating functions for the state and operator can be unified in a simple way.

\section{Numerical results}

\subsection{Variational excited states and dynamical structure factor}

\begin{figure}[!hbt]
\centering
	\includegraphics[width=0.95\columnwidth]{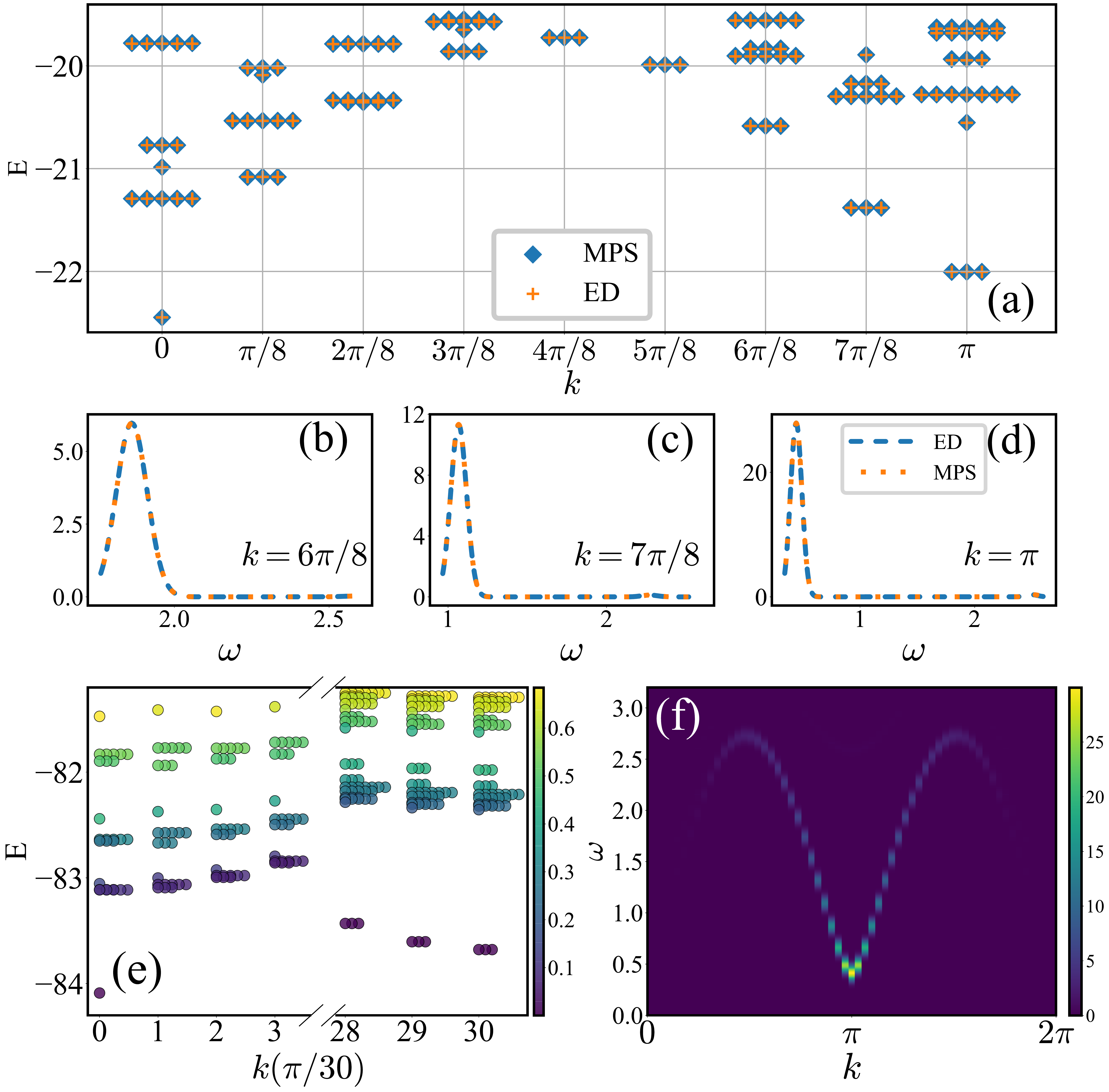}
\caption{Spectral properties of spin-$1$ Heisenberg chain. (a) shows a comparison of $D=24$ variational MPS calculation with ED for system size $N=16$. Further comparison of DSF at $k=3\pi/4, 7\pi/8, \pi$ is shown in (b)-(d). In (e), we show the low-energy spectrum with size $N=60$, obtained with $D=30$ MPS, where color represents the energy variance. Based on the spectrum in (e), DSF is computed and shown in (f) with a clear peak around $(k, \omega)=(\pi, 0.4)$.}
\label{fig:HeisenbergSpec}
\end{figure}

We now present the numerical results using generating functions. The model we use for benchmark is the spin-$1$ Heisenberg chain with Hamiltonian: $\hat{H}=J\sum_{j=1}^{N}\mathbf{S}_j\cdot\mathbf{S}_{j+1}$, where $\mathbf{S}_j=(\hat{S}_j^x,\hat{S}_j^y,\hat{S}_j^z)$ are spin-1 operators on site-$j$ and $J=1$ is taken as the energy unit. We use periodic uniform MPS to approximate the ground state, and compute the low-energy spectrum with excitation ansatz, where generating functions introduced above are used. For small system size ($N=16$), we compare the variational energy spectrum with that obtained from exact diagonalization (ED). As shown in Fig.~\ref{fig:HeisenbergSpec}(a), with bond dimension $D=24$, the variational result agrees remarkably well with ED, with maximal relative deviation $6\times 10^{-4}$ for high energy levels, and the level degeneracy is also recovered to a high precision. Moreover, although the low-energy excitations around $k=0$ are two-magnon continuum~\cite{Affleck1992, White2008}, it is nevertheless reproduced well by one-particle excitation ansatz.

With excited states available, dynamical properties of the system can be investigated in a straightforward manner by constraining the total Hilbert space to subspace spanned by variational ground and excited states~\cite{Ferrari2019}. One of the key observables is the dynamical spin structure factor (DSF), which reveals properties of quasiparticles and is accessible in neutron scattering experiments. The DSF is defined as: $S^{\alpha}(k,\omega)=\sum_{n}|M^{\alpha}_k|^2\delta(\omega-E^k_n+E_0)$, with $M^{\alpha}_k=|\langle\Phi_k(B_n)|\hat{S}^{\alpha}_k|\Psi(A)\rangle|$ and $E_0$ ($E^k_n$) being energy of ground state $|\Psi(A)\rangle$ ($n$-th excited state $|\Phi_k(B_n)\rangle$ with momentum $k$). The delta function is replaced by a normalized Gaussian with broadening width $\sigma=0.05$, and we further take $\hat{S}^z$ as the operator in DSF. The DSF can be efficiently computed with generating functions Eqs.~\eqref{eq:excitationGen}, \eqref{eq:sfGen}, and the results for $k=3\pi/4, 7\pi/8, \pi$ are shown in Fig.~\ref{fig:HeisenbergSpec}(b)-(d) with system size $N=16$. Comparing with ED, the line shapes are reproduced well by excitation ansatz.

For larger system size ($N=60$), we quantify the quality of the variational result with the energy variance $\mathrm{Var}(E)=\langle \hat{H}^2\rangle-\langle \hat{H} \rangle^2$, computed using generating functions (see Eqs.~\eqref{eq:normGen}, \eqref{eq:hamGen}). Here the bond dimension is $D=30$. As shown in Fig.~\ref{fig:HeisenbergSpec}(e), the energy variance of the variational ground state and one-magnon excited states remains small (on the order of $1\times 10^{-3}$), considering the large system size, while the variational multi-magnon excitations appear to have larger energy variances. The Haldane gap can be read off as $\Delta=0.4105$, in agreement with Refs.~\cite{White2008, Haegeman2012}. Further evaluating DSF using excited states (shown in Fig.~\ref{fig:HeisenbergSpec}(f)), we find a strong peak appearing at $(k,\omega)=(\pi,\Delta)$, where the first excited state is located. This confirms that the elementary excitation is magnon with momentum $k=\pi$, in agreement with the variational spectrum where the first excited state is a triplet at $k=\pi$. Vanishing DSF around $k=0$ is also consistent with excitations being two-magnon continuum in that region~\cite{White2008}.

\subsection{R\'enyi entropy of excited states}

Apart from variational energy spectrum and DSF, the generating functions also allow us to further study properties of excited states in great detail. Here we use them to investigate entanglement properties of excited states, e.g., R\'enyi entropy, which has received considerable interest in recent study~\cite{Alcaraz2011, Molter2014, Castro2018, Zhang2020}. It is well known that for 1D gapped systems, the entanglement entropy of ground state saturates with increasing subsystem size~\cite{Eisert2010}. However, much less is known for excited states. Traditionally, computing R\'enyi entropy for excited states requires multiple summations and thus is hard to achieve. Here we explore this question using generating functions without any summations.

For a normalized excited state $|\Phi\rangle$ with bipartition of the system $L$ and $\overline{L}$, reduced density matrix (RDM) of subsystem $L$ (with size $l$) is given by $\rho_L=\mathrm{Tr}_{\overline{L}}|\Phi\rangle\langle\Phi|$, and the R\'enyi entropy is then defined as $S^{(n)}=\frac{1}{1-n}\mathrm{ln}\mathrm{Tr}_L\rho_L^n$. Here we will focus on the $n=2$ case, which can be computed by taking two copies of $|G_{\Phi}(\lambda)\rangle$ and $\langle G_{\Phi}(\lambda)|$, each with an independent parameter $\lambda$. Through AD of a single diagram, a fourth order derivative (one for each layer) at zero point gives directly the R\'enyi entropy.

\begin{figure}[!t]
\centering
	\includegraphics[width=0.95\columnwidth]{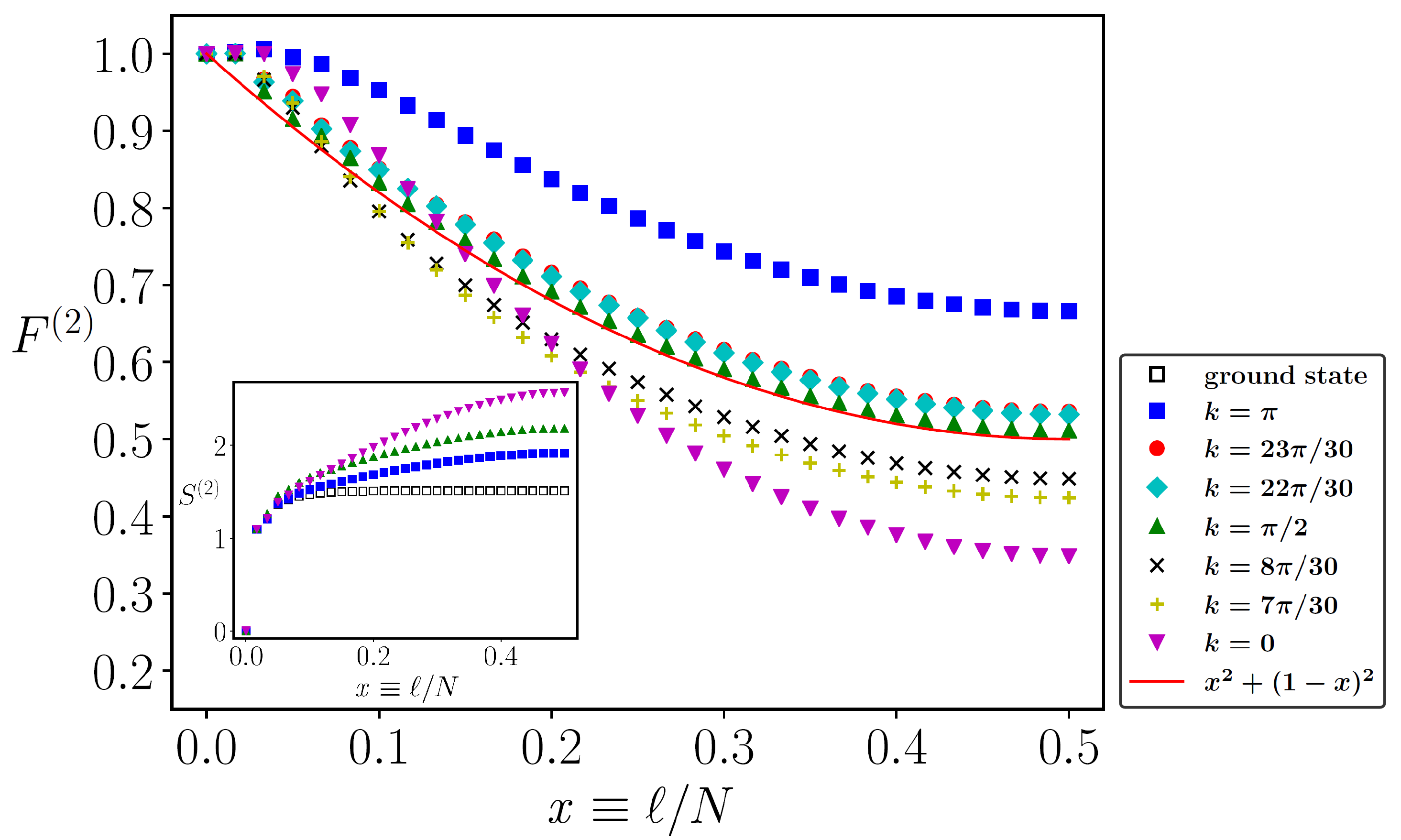}
\caption{R\'enyi entropy of excited states for spin-1 Heisenberg model. In the main panel, different symbols represent the quantity $F^{(2)}$ of first excited state at various $k$ with subsystem size $x\equiv l/N$. Red line shows the theoretical prediction. Typical R\'enyi entropy with subsystem size is shown in the inset.}
\label{fig:HeisenbergEntropy}
\end{figure}

The R\'enyi entropy with $n=2$ for spin-$1$ Heisenberg model is shown in Fig.~\ref{fig:HeisenbergEntropy}, using excited states in Fig.~\ref{fig:HeisenbergSpec}(e). The saturation of ground state R\'enyi entropy with increasing $l$ is evident in the inset of Fig.~\ref{fig:HeisenbergEntropy}, as expected. However, for excited states, we find that generically the R\'enyi entropy does not saturate with increasing $l$. To further quantify the effect of quasiparticles in excited states, we consider the ratio between excited and ground state traces: $F^{(2)}=\frac{\mathrm{Tr}\rho_{\Phi}^2}{\mathrm{Tr}\rho_{\Psi}^2}$~\cite{Alcaraz2011}, where $\rho_{\Phi} (\rho_{\Psi})$ is RDM of excited (ground) state with subsystem size $l$. Theoretical studies have shown that, under the assumption of large momentum and energy gap, $F^{(2)}$ takes a universal form~\cite{Castro2018}: $F^{(2)}=x^2+(1-x)^2$, with $x\equiv l/N$. Indeed, the $k=\pi/2$ result agrees with the theoretical prediction, although our model is neither integrable nor free. However, clear deviation is also observed for other momenta, which could be ascribed to the small energy gap or momentum and deserves further study.

\section{Discussion}
 
Above we have shown that the generating functions can be used to compute one-particle excitation spectrum of a quantum spin chain with finite length. The results are encouraging, as the low-energy content is fully captured by MPS excitations, finding direct applications to investigate edge properties of two dimensional (2D) system~\cite{Cirac2020}. The generating functions can also be adapted for tangent space based excitation ansatz without translation symmetry~\cite{VanDamme2021}. There are several directions for further extension. To go beyond one-particle case, one option would be modifying the ground state tensor $A$ with $A+\sum_i\lambda_iB_i$, similar to introducing several source fields in partition function of QFT. For infinite size systems, one can combine generating functions with fixed-point methods, which will enable computing excitation spectrum of 2D system with projected entangled-pair states, including anyonic excitations in topological phases of matter~\cite{Cirac2020}.

\section{Conclusion and outlook}

In this work, we have introduced a set of generating functions for both TN states and operators, thereby eliminating extensive diagrammatic summations in modern applications of TN methods. Using generating functions, we have shown that excitation spectrum of quantum spin chain can be computed efficiently and accurately with periodic uniform MPS, and the procedure is formally the same as ground state search using TNs. Moreover, the generating functions allow us to investigate dynamical structure factor of the system and entanglement property of excited states in a convenient way, the later of which is beyond the capability of traditional methods. We envision the generating functions introduced here will be powerful in the next generation of tensor network algorithms and applications.

{\it Note added} -- During the preparation of this manuscript, we became aware of an independent work by Boris Ponsioen and Philippe Corboz~\footnote{B. Ponsioen and P. Corboz (unpublished).}.

\section*{Acknowledgements}

We thank useful conversations with Hyun-Yong Lee, Fa-Hui Lin, Cosimo C. Rusconi, Hong-Hao Tu, and Jiaju Zhang. Part of the numerical calculation was performed at the Supercomputer Center, ISSP, University of Tokyo. WLT and NK are supported by the KAKENHI, Project ID: MEXT Grant-in-Aid for Scientific Research (B)  (19H01809). HKW is supported by JQI-NSF-PFC (NSF grant PHY-1607611). JYC and NS acknowledge support by the European Union's Horizon 2020 programme through the ERC Starting Grant WASCOSYS (Grant No.~636201) and the ERC Consolidator Grant SEQUAM (Grant No.~863476), and from the DFG (German Research Foundation) under Germany's Excellence Strategy (EXC-2111 -- 390814868).

\bibliography{bibliography}

\end{document}